\documentclass[pra,twocolumn,showpacs,superscriptaddress,amsfonts,amsmath,floatfix]{revtex4-1}

\usepackage{graphicx,amsfonts}

\begin{document}

\title{Probing superfluidity of a mesoscopic Tonks-Girardeau gas} 
\author{C. Schenke} 
\email{Christoph.Schenke@grenoble.cnrs.fr} 
\affiliation{Universit\'e Grenoble 1/CNRS, Laboratoire de Physique et de Mod\'elisation des Milieux Condens\'es, UMR5493,  B.P. 166, 38042 Grenoble, France} 
\author{A. Minguzzi} 
\affiliation{Universit\'e Grenoble 1/CNRS, Laboratoire de Physique et de Mod\'elisation des Milieux Condens\'es, UMR5493,  B.P. 166, 38042 Grenoble, France} 
\author{F.W.J. Hekking} 
\affiliation{Universit\'e Grenoble 1/CNRS, Laboratoire de Physique et de Mod\'elisation des Milieux Condens\'es, UMR5493,  B.P. 166, 38042 Grenoble, France}

\begin{abstract}
We study the dynamical response of a Tonks-Girardeau gas on a ring induced by a  moving delta-barrier potential. An exact solution based on the
time-dependent Bose-Fermi mapping allows to obtain the particle current, its fluctuations and the drag force acting on the barrier. The exact solution is analyzed numerically as well as analytically in the perturbative regime of weak barrier strength. In the weak barrier limit the stirring drives the system into a state with net zero current for velocities $v$ smaller than $v_c=\pi\hbar /mL$, with $m$ the atomic mass and $L$ the ring circumference. At $v$ approaching $v_c$ angular momentum can be transferred to the fluid and a nonzero drag force arises. The existence of a velocity threshold for current generation indicates superfluid-like behavior of the mesoscopic Tonks-Girardeau gas,  different from the non-superfluid  behavior predicted for Tonks-Girardeau gas in an infinite tube.  
\end{abstract}
\maketitle

\section{Introduction}
A typical superfluid displays frictionless flow past an obstacle. Superfluid behavior occurs at small flow velocities and breaks down at a critical velocity threshold $v_c$, where  creation of elementary excitations becomes energetically possible  \cite{Landau_book}. This phenomenon can be viewed in the co-moving reference frame, where impurities or tube walls move with respect to the fluid. Following the same idea it is possible to probe  superfluid-like behavior by driving a moving barrier with respect to a fluid at rest. This ``stirring'' perturbation has been applied to probe superfluidity in a three-dimensional elongated Bose-Einstein condensate by moving a blue-detuned laser beam inside the condensate \cite{stirring_raman_99}.  

For a one-dimensional (1D) fluid, the presence of a barrier affects the fluid particularly strongly as particles cannot circumvent it; moreover in 1D quantum fluctuations are important. Both aspects are expected to affect the transport behavior of a 1D fluid induced by a moving barrier with velocity $v$. The possibility of observing superfluid-like behavior in 1D has been addressed by Astrakharchik and Pitaevskii  \cite{astrakharchik_stirring}. They have shown that for an infinite tube there is no finite critical velocity, but there is a non vanishing drag force $F_{d}\propto v^{2K-1}$, where $K$ is the Luttinger parameter. For a weakly interacting 1D Bose gas $K\gg 1$ and the drag force at small velocities is very small: the system behaves almost like a 3D superfluid. 
In this limit the drag force has been calculated using the  Bogoliubov approximation \cite{Syk09}. 
On the other hand, in the Tonks-Girardeau (TG) limit of impenetrable bosons \cite{Gir1960,Gir1965} one finds $K=1$ (see eg \cite{Cazalilla03}) and the drag force scales linearly with the velocity as in a normal fluid. Calculations of the drag force for a 1D Bose gas at large interaction strengths are given in \cite{CheCauBra09}. The link between drag force and 1D superfluidity has been reviewed in \cite{CheCauBra11}.

The peculiar  transport properties of 1D  fluids have been widely explored in the case of electrons. For an infinite fermionic Luttinger liquid the transport behavior across a barrier depends dramatically on the strength of the Luttinger parameter \cite{kane_fisher_prl}. For $K<1$ (eg repulsive interactions between fermions) the particles are perfectly reflected even by a very small barrier -- leading to an insulating behavior, while for $K>1$ they are perfectly transmitted even by a very large barrier -- displaying superfluid-like behavior. This change of behavior as $K$ is increased through $K=1$ is also known as the Schmid transition \cite{Schmid}. The case $K=1$ (noninteracting fermions) is exactly solvable: for an infinite wire the electronic conductance across a large  barrier is small \cite{kane_fisher_prl}, hence its behavior is closer to the one of an insulator. A similar analysis can be done in the bosonic case \cite{buchler}, where the Bose gas with short-range interactions ($K>1$) on an infinite tube is predicted to show a superfluid-like behavior at zero temperature, and to undergo a transition to a non-superfluid state at $K=1$, i.e. the TG limit. This transition  \cite{Schmid} can be viewed as a localization-delocalization transition of the relative phase across the barrier: a superfluid has a well-defined, hence localized relative phase, while the insulator has a fluctuating, delocalized phase. Superfluidity is destroyed by quantum or thermal fluctuations which induce phase slips, i.e. jumps of the phase between potential-energy minima.

In the case of a finite ring of circumference $L$  the transport properties of a 1D bosonic fluid  -- as probed by a moving barrier potential -- change considerably at small velocities. A Luttinger-liquid approach to a weakly interacting Bose gas  \cite{buchler} predicts a superfluid-like behavior  with critical velocity $v_c=\pi \hbar/mL$, $m$ being the mass of the bosons. The question naturally arises as to what happens for strong interactions, especially for $K=1$ where, according to the analysis of the drag force,  the infinite system displays nonsuperfluid behavior.   

To answer this question we explore the dynamical properties of  a mesoscopic TG gas, stirred by a moving delta barrier set into motion either abruptly or adiabatically. This model is of interest since ultracold atomic gases trapped in tight ring traps are currently being experimentally explored \cite{Gup05,ArnGarRii06,Mor06,Hen09,Ram11,She11,Mou11} and the TG gas has been experimentally observed in finite 1D tubes \cite{Kino04,Pare04}. Using an exact mapping onto a noninteracting Fermi gas \cite{Gir1960,GirWri00}, we can access the exact many-body wavefunction, allowing us to analyze the details of its spatial structure and the whole dynamical evolution. By calculating the integrated current induced in the fluid, the current fluctuations and the drag force acting on the barrier, as a main result we find for the TG gas a superfluid-like behavior with the same critical velocity $v_c$ as for the weakly interacting Bose gas. 

\section{Dynamical evolution under stirring drive}
The Hamiltonian of  $N$ bosons on a ring of circumference $L$, stirred by a  delta potential $U(x,t)=U_0 \delta(x-vt)$ is given by 
\begin{equation}
\hat{H}_B=\sum_{j=1}^{N}\left[-\frac{\hbar^2}{2m}\frac{\partial^2}{\partial x_{j}^2}
+U(x_{j},t)\right]+\sum_{j<\ell}g \delta (x_j-x_\ell).
\end{equation}
In the TG regime,  corresponding to the impenetrable boson limit  $g\to\infty$,  the interaction potential can be replaced by a cusp  condition on the  many-body wavefunction
\begin{equation}
\label{eq:cusp}
 \Psi_B(...x_j=x_\ell...)=0. 
\end{equation}
For a gas on a ring we further impose periodic boundary conditions $\Psi_B(...x_j...)=\Psi_B(...x_j+L...)$ for any $j$. 

It is possible to obtain an exact analytical solution for the many-body wavefunction by mapping the system onto a gas of noninteracting fermions subjected to the same external (time-dependent) potential \cite{GirWri00},
\begin{equation}
\label{eq:manybody}
\Psi_B(x_1,...x_N,t)= {\cal A}(x_1,...x_N) (1/\sqrt{N!})\det[\psi_l(x_m,t)]
\end{equation}
with ${\cal A}(x_1,...x_N) =\Pi_{j<\ell}{\rm sign}(x_j-x_\ell)$ being a mapping function which ensures the  bosonic symmetry under exchange of two particles.
In each coordinate sector $x_{P(1)}<x_{P(2)}<...<x_{P(N)}$, where $P$ is a permutation of the set $\{1,2,...N\}$,  the many-body wavefunction Eq.(\ref{eq:manybody}) satisfies  the required boundary conditions Eq.(\ref{eq:cusp}) and is the unique solution of  the many-body time-dependent Schroedinger equation, provided that the orbitals $ \psi_l(x,t)$ satisfy  the single-particle  Schroedinger equation
\begin{equation}
\label{tdstirring}
i\hbar \partial_t \psi_l(x,t)=\left(-\frac{\hbar^2}{2m}  \partial_x^2 + U_0 \delta(x-vt)\right) \psi_l(x,t).
\end{equation}

\subsection{Exact Solution}

\begin{figure}
\centerline{\includegraphics[width=0.4\textwidth]{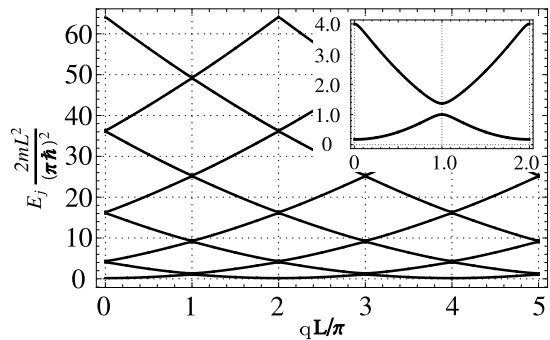}} 
\caption{The energy eigenvalues in units of $(\pi\hbar)^2/(2mL^2)$ as a function of the stirring wavevector $q$ in units of $\pi/L$. Each parabola represents a state of different angular momentum. These states are coupled by the barrier that opens gaps at $q_n=n\pi/L$ and thus allows to change the branch of angular momentum. The inset shows the first avoided level crossing between the states with zero and one quantum of angular momentum.
}
\label{fig1}
\end{figure}

If the barrier is set into motion instantaneously at $t=0$ we have found an analytical solution for Eq.(\ref{tdstirring}) \cite {SchMinHek11}.
We choose as the initial condition the ground state orbitals of a TG gas  in the presence of a nonmoving barrier localized at $x=0$  \cite{GirMing09}, i.e. in Eq.(\ref{eq:manybody}) we set  $\psi_l(x,0)=\phi^{(0)}_l(x)$, where  for spatially even orbitals we have 
\begin{equation}
\label{eq:even_orb}
\phi^{(0)}_l(x)=\frac{2}{{\cal N}_l^{(0)}} \cos[k_l^{(0)} (|x|-L/2)]
\end{equation}
with wavevectors fixed by the transcendental equation $k_l^{(0)} \tan (k_l^{(0)}L/2)=mU_0/\hbar^2$ and normalization given by ${\cal N}_l^{(0)}=\sqrt{2L(1+\sin(k_l^{(0)}L)/k_l^{(0)} L)}$, while for odd orbitals we have
\begin{equation}
\label{eq:odd_orb}
\phi^{(0)}_l(x)=i (-1)^l\sqrt{\frac{2}{L}} \sin[2 \pi l x/L],
\end{equation}
with $l=1,2,3...$ integer, and the phase factor in Eq.(\ref{eq:odd_orb}) has been chosen for consistency with the forthcoming Eq.(\ref{eq:phij}).

The solution for the dynamical evolution according to Eq.(\ref{tdstirring}) then  reads
\begin{equation}
\label{eq:orbitals}
\psi_l(x,t)=e^{i qx} e^{-iq^2t/2m}\sum_j c_{jl}e^{-i E_{j,q} t/\hbar}\phi_{j,q}(x-vt),
\end{equation}
where $\hbar q=mv$ is the barrier momentum. The time-independent overlaps $c_{jl}$ contain the information about the initial condition and are defined as
\begin{equation}
\label{eq:coeff}
c_{jl}=\langle \phi_{j,q}|e^{-iqx}|\phi^{(0)}_l\rangle. 
\end{equation}
The orbitals $ \phi_{j,q}(x)$  and the energies $E_{j,q}$ are obtained from the stationary Schroedinger equation in the frame co-moving with the barrier,
\begin{equation}  
\label{eq:defphij}
\left(-\frac{\hbar^2}{2m}  \partial_x^2 + U_0 \delta(x)\right)  \phi_{j,q}=E_{j,q} \phi_{j,q}
\end{equation}
with twisted boundary conditions (TBCs) $\phi_{j,q}(x+L)=e^{-iqL} \phi_{j,q}(x)$. The solution of Eq.(\ref{eq:defphij}) is obtained by expansion with respect to  plane waves, 
\begin{equation}
\label{eq:phij}
\phi_{j,q}(x)=\frac{1}{{\cal N}_j}\! 
\begin{cases}
\!e^{iq\frac{L}{2}}(e^{i(k_j(x+\frac{L}{2})}\!+\!A_{j} e^{-i k_j(x+\frac{L}{2})}) \!\!\!\!\!\!\!\!
&\in  [-\frac{L}{2},0) \\
\!e^{-iq\frac{L}{2}}(e^{i(k_j(x-\frac{L}{2})}\!+\!A_{j} e^{-i k_j(x-\frac{L}{2})}) \!\!\!\!\!\! & \in  [0,\frac{L}{2}]. \end{cases} 
\end{equation}
The  normalization factor ${\cal N}_j=\sqrt{L(1+ A_j^2+ 2 A_j \sin(k_jL)/k_jL})$  and the  amplitude $A_j=\sin[(k_j+q)L/2]/\sin[(k_j-q)L/2]$ are functions of the  wavevectors $k_j$ given by the solution of the transcendental equation 
\begin{equation}
\label{eq:transc}
k_j=\frac{mU_0}{\hbar^2} \frac{\sin(k_jL)}{\cos(qL)-\cos(k_jL)}.
\end{equation}
This also fixes the energy eigenvalues  $E_j=\hbar^2 k_j^2/2m$ that are shown in Fig.\ref{fig1}. Finally, using Eq.(\ref{eq:phij}) and respectively Eqs.(\ref{eq:even_orb}) and (\ref{eq:odd_orb}) above, it is possible to find an analytical expression for the overlap coefficients, which read
\begin{small}
\begin{eqnarray}
c_{jl}^{e}&=&\frac{L}{{\cal N}_j{\cal N}^{(0)}_l }\left\{\!J_0\!\left[(k_l^{(0)}\!-k_j-q)\frac{L}{2}\right]\!+\!J_0\!\left[(k_l^{(0)}\!+k_j+q)\frac{L}{2}\right]\right.\nonumber \\
&+&\!\!\left. A_j J_0\!\left[(k_l^{(0)}\!+k_j-q)\frac{L}{2}\right]\!+\!A_j J_0\!\left[(k_l^{(0)}\!-k_j+q)\frac{L}{2}\right] \!\right\}
\end{eqnarray}
\end{small}
\hspace{-0.1cm}for the overlap with an initial even orbital, and
\begin{small}
\begin{eqnarray}
c_{jl}^{o}&=&\frac{\sqrt{L/2}}{{\cal N}_j}\!\left\{\!J_0\!\left[(k_l^{(0)}\!-k_j-q)\frac{L}{2}\right]\!-\!J_0\!\left[(k_l^{(0)}\!+k_j+q)\frac{L}{2}\right]\right.\nonumber \\
&+&\!\!\left. A_j J_0\!\left[(k_l^{(0)}\!+k_j-q)\frac{L}{2}\right]\!-\!A_j J_0\!\left[(k_l^{(0)}\!-k_j+q)\frac{L}{2}\right] \!\right\}
\end{eqnarray}
\end{small}
\hspace{-0.1cm}for the overlap with an initial odd orbital,
with $J_0(x)=\sin(x)/x$ being the first spherical Bessel function.

\begin{figure}
\centerline{\includegraphics[width=0.4\textwidth]{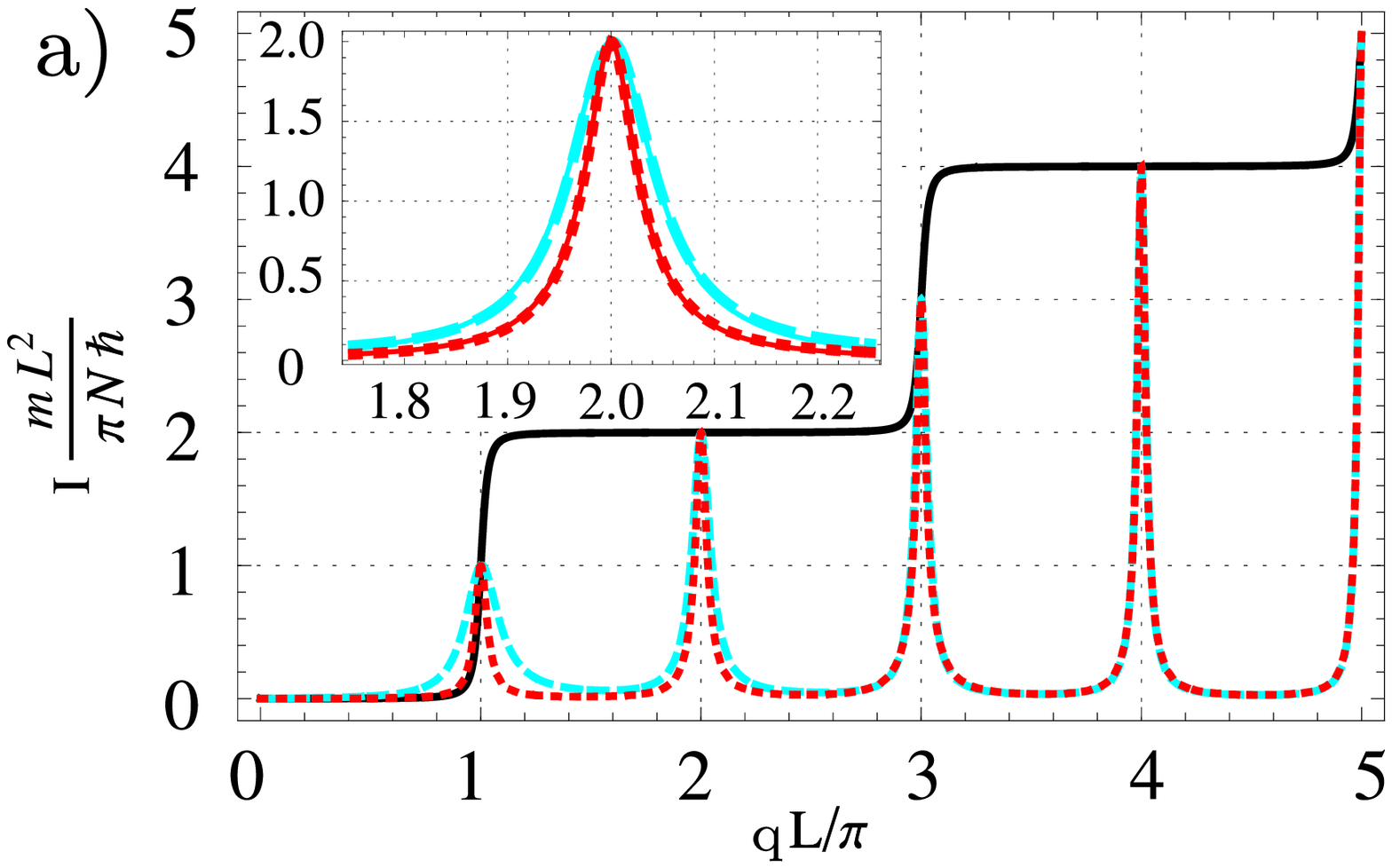}} \centerline{\includegraphics[width=0.4\textwidth]{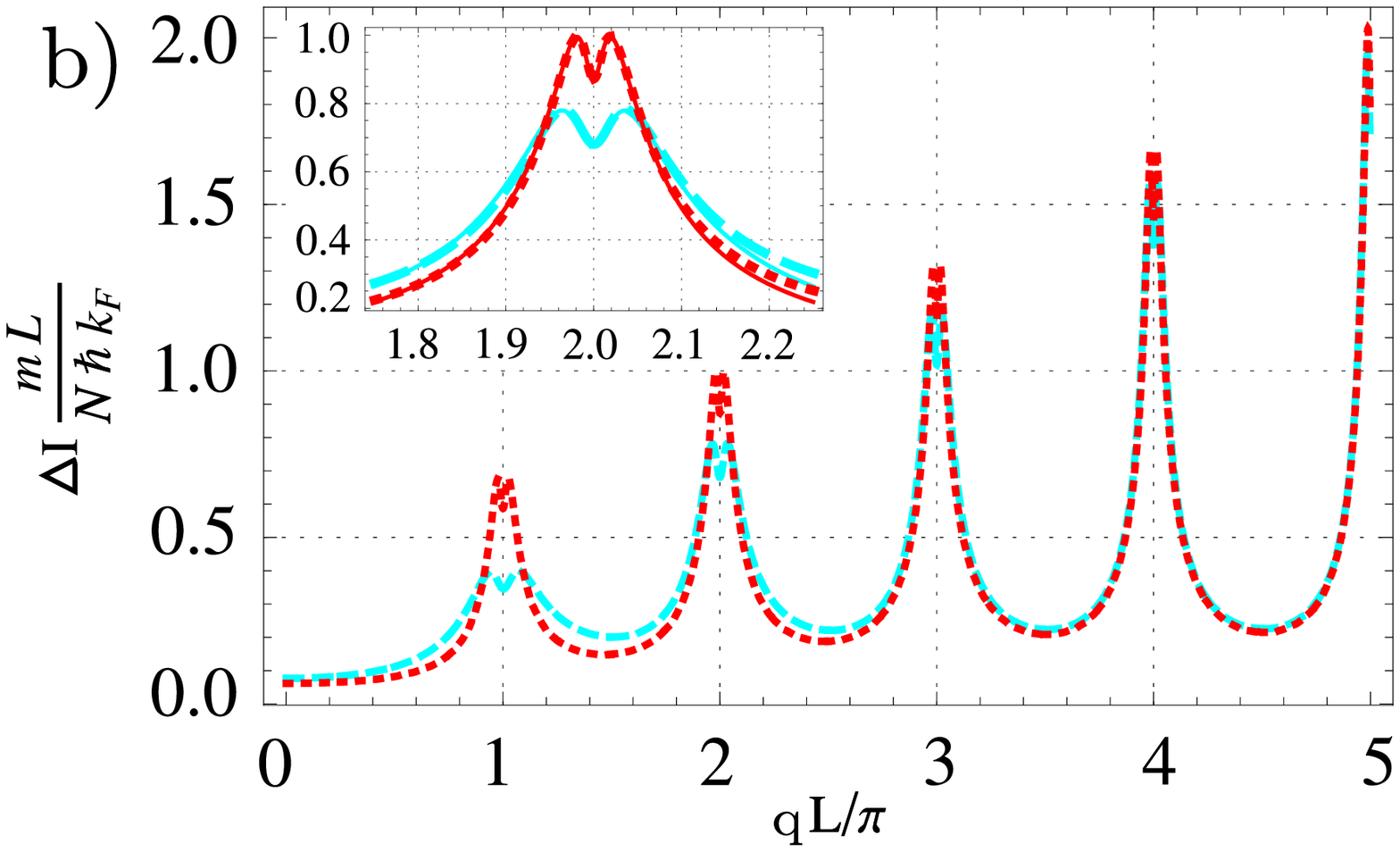}} 
\caption{(Color online) a) Time-averaged, integrated particle current in units of $(\pi N\hbar)/(mL^2)$ and b) current fluctuations in units of $(N\hbar k_F)/(mL)$, $k_F$ being the initial wavevector of the $Nth$ particle, for a TG gas (red dotted line) and a noninteracting Bose gas (cyan dashed line) subjected to the nonadiabatic stirring described in the text, as a function of the stirring wavevector $q$ in units of $\pi/L$, for a barrier strength $m U_0L/\hbar^2=1$ and $N=3$ particles. The behavior of the TG current following  an adiabatic switching on is also shown (black solid line). Insets: Zoom on the second current (current fluctuation) peak. The numerical results (same notation and color codes as the main panel) are compared with the results of the perturbative approach (thin solid lines).
}
\label{fig2}
\end{figure}

\subsection{Perturbative solution for weak barriers}
In order to develop a qualitative understanding of the behavior of the system under stirring, we study the case of weak barriers, i.e. for $\lambda L \lesssim 1$, where it is possible to obtain the dynamical evolution of the wavefunction (\ref{eq:orbitals}) perturbatively. 
Due to the ring periodicity the problem of the time-independent Hamiltonian in Eq.(\ref{eq:defphij}) with solutions that obey TBCs is equivalent to the one for a particle in a periodic potential and we can therefore expand its solutions in Bloch waves
\begin{equation}  
\label{eq:BlWav}
\phi_{j,q}(x)=\frac{e^{-iqx}}{\sqrt{L}}\sum_{p\in \mathbb{Z}} \alpha_{q-2\pi p/L}^{(j)}e^{i2\pi px/L} ,
\end{equation} 
with eigenvalues $E_{j,q}$ and coefficients $\alpha_{q+2\pi p/L}^{(j)}$, $j$ being the band index. Indeed the eigenstates $\phi_{j,q}(x)$ of the Hamiltonian in Eq.(\ref{eq:defphij}) are written as a product of a plane wave $\exp(-iqx)$ and a lattice- (in our case ring-) periodic function and thus obey Bloch's theorem. The parameter $q$ plays the role of the quasi-momentum. It provides a periodic continuous parametric dependence of the states and the energy on the stirring velocity. If the barrier is weak it will only couple two states $2\pi a/L$ and $2\pi b/L$, $a,b\in\mathbb{Z}$ of different angular momentum depending on the concerned avoided level-crossing, see Fig.\ref{fig1}. Consequently, at fixed avoided level crossings around $q_n=n \pi/L$, $n$ being an odd integer only two elements of the sum in Eq.(\ref{eq:BlWav}) contribute and it reduces to
\begin{eqnarray}
\label{eq:BlWavPT1}
\phi^{(pert)}_{j,q}(x)\!&=&\!\frac{e^{-iqx}}{\sqrt{L}}\!\left[\alpha_{q-\frac{\pi}{L} (n-j)}^{(j)}e^{i(\frac{\pi}{L} (n-j)x}\right.\nonumber\\
&&\qquad\qquad\qquad\left.+\alpha_{q-\frac{\pi}{L} (n+j)}^{(j)}e^{i(\frac{\pi}{L} (n+j)x}\right] \\
\label{eq:BlWavPT2}
\phi^{(pert)}_{j+1,q}(x)\!&=&\!\frac{e^{-iqx}}{\sqrt{L}}\!\left[\alpha_{q-\frac{\pi}{L} (n-j)}^{(j+1)}e^{i(\frac{\pi}{L} (n-j)x}\right.\nonumber\\
&&\qquad\qquad\qquad\left.+\alpha_{q-\frac{\pi}{L} (n+j)}^{(j+1)}e^{i(\frac{\pi}{L} (n+j)x}\right]\quad\
\end{eqnarray}
for odd values of $j$. The expansion coefficients $\alpha_{q-2\pi p/L}^{(j)}$ are readily obtained by solution of the effective two-level problem,
\begin{eqnarray}
\alpha_{q-\frac{\pi}{L} (n-j)}^{(j)}&=&\alpha_{q-\frac{\pi}{L} (n+j)}^{(j+1)}=v_q^{(j)}\\
\alpha_{q-\frac{\pi}{L} (n+j)}^{(j)}&=&-\alpha_{q-\frac{\pi}{L} (n-j)}^{(j+1)}=-u_q^{(j)}
\end{eqnarray}
where
\begin{eqnarray}
\left\{u_q^{(j)},v_q^{(j)}\right\}=
\left[\frac{1}{2}\pm\frac{\frac{\hbar^2}{m}\ \frac{j\pi}{L}\ \delta q}{\sqrt{\left(\frac{2\hbar^2}{m}\ \frac{j\pi}{L}\ \delta q\right)^2+\left(\frac{2U_0}{L}\right)^2}}\right]^{1/2}\! ,\qquad 
\end{eqnarray}
the +(-) solution referring to $u_q^{(j)}$ ($v_q^{(j)}$), and we have introduced $\delta q=q-q_n$, the deviation of $q$ from $q_n$. Equations (\ref{eq:BlWavPT1}) and (\ref{eq:BlWavPT2}) also hold for even values of $n$ choosing $j$ even \cite{note}.

We finally obtain the full wavefunction at the avoided level crossing inserting Eq.(\ref{eq:BlWavPT1}) and (\ref{eq:BlWavPT2}) in Eq.(\ref{eq:orbitals}), 
\begin{multline}  
\label{eq:WavPT}
\psi^{(pert)}_{l}(x,t)=\\e^{iqx}e^{-iq^2t/2m}\sum_j\! '\left(c_{jl}\ e^{iE_{j,q}t/\hbar}\ \phi^{(pert)}_{j,q}(x-vt)\right.\\ \left.+c_{j+1,l}\ e^{iE_{j+1,q}t/\hbar}\ \phi^{(pert)}_{j+1,q}(x-vt)\right)\ ,
\end{multline} 
with 
\begin{multline}  
\label{eq:EnergPT}
\left\{E_{j,q},E_{j+1,q}\right\}=\frac{\hbar^2}{2m}\left[\left(\delta q\right)^2+\left(\frac{j\pi}{L}\right)^2\right]\\
\mp\sqrt{\left(\frac{\hbar^2}{m}\ \frac{j\pi}{L}\ \delta q\right)^2+\left(\frac{U_0}{L}\right)^2}
\end{multline} 
being the energy of the $j$th and the $(j+1)$th band. Here the $\sum'$ is over the even $j$ or the odd $j$ only, depending on $n$. The coefficients $c_{jl}$ are determined using Eq.(\ref{eq:coeff}) where the initial orbitals are plane waves $\phi^{(in)}_{l}=\exp(ik_lx)/\sqrt{L}$ with $k_{l}=-\pi(l-1)/L$ for odd $l$ and $k_{l}=\pi l/L$ for even $l$, yielding
\begin{eqnarray}
\begin{array}{cc}
\begin{array}{c}\quad\ \ c_{jl}=v_q^{(j)}\delta_{j,n+l-1}\\ \quad c_{j+1,l}=u_q^{(j)}\delta_{j,n+l-1} \end{array}& \ \mbox{for}\quad l\ \mbox{odd}\\\\
\begin{array}{c}\ \ c_{jl}=v_q^{(j)}\delta_{j,n-l}\\c_{j+1,l}=u_q^{(j)}\delta_{j,n-l} \end{array}& \ \mbox{for}\quad l\ \mbox{even}
\end{array}\ .
\end{eqnarray}

\section{Integrated particle current}
The knowledge of the many-body wavefunction allows to derive several observables of interest. We focus first on  the integrated particle current 
$I=(1/L)\int dx j(x,t)$, which for the ring geometry is proportional to the angular momentum.  The current-density $j(x,t)$ coincides with the one of the mapped Fermi gas \cite{Citro09},  
\begin{equation}
j(x,t) =\frac{\hbar}{ m} {\rm Im} \sum_{l=1}^N\psi_l^*(x,t)\partial_x \psi_l(x,t).
\end{equation}
Using this expression we evaluate the integrated current
\begin{eqnarray}
 \label{eq:int_current}
I(t)= \frac{N\hbar}{mL}q+\frac{\hbar}{mL} \sum_{l=1}^N \left\{  -\sum_j |c_jl|^2  \frac{\partial}{\partial q} \left(\frac{k_j^2}{2}\right)\right. \nonumber \\
+ \left.{\rm Im} \sum_{i\neq j}c^*_{il} c_{jl} e^{-i(E_j-E_i)t/\hbar}F_{ij} \right\},
\end{eqnarray}
where the amplitudes $F_{ij}=\langle \phi_i| \partial_x| \phi_j\rangle$ are given by
\begin{multline}
F_{ij}=\frac{-i k_ik_j L^2}{{\cal N}_i{\cal N}_j} J_0[(k_i-k_j)L/2] J_0[(k_i+k_j)L/2]\\ \times \frac{\sin(qL)}{\sin[(k_i-q)L/2]\sin[(k_j-q)L/2]}\ .
\end{multline}
The result (\ref{eq:int_current}) for the spatially integrated current  displays a time-independent contribution which has the form of a generalized thermodynamic relation $I\propto\partial E/\partial q$ and a time-dependent oscillating term, with several frequency components originating from the multimode nature of the state of the TG gas.
In order to gain further insight we focus on the case of weak barrier strength. Close to any level crossing, Eq.(\ref{eq:WavPT}) allows to obtain an explicit expression for the integrated particle current,
\begin{multline}
 \label{eq:int_currentPT}
I(t)= \frac{\hbar}{mL}\!\sum_{l=1}^N 4(q_n-k_l)u_q^{(f_{nl})}v_q^{(f_{nl})}\\
\times\left[1-\cos(2\Delta E_q^{(f_{nl})}t)\right] ,
\end{multline}
where we introduced
\begin{eqnarray}
f_{nl}=
\left\{
\begin{array}{cc}
 |n-l| & \ \mbox{for}\quad l\ \mbox{even}\\
 (n+l-1) & \ \mbox{for}\quad l\ \mbox{odd}
\end{array}
\right.\ .
\end{eqnarray}
Here $\Delta E_q^{(l)}=E_{l+1,q}-E_{l,q}$ is the energy splitting of the effective two-level system that governs the time-dependence of the current. If averaged over sufficiently long times $T\gtrsim\hbar/\Delta E_q$, this dependence vanishes.
In Fig.\ref{fig2}a we illustrate the resulting time-averaged current for a weak barrier. We find that it displays a narrow maximum for values of the stirring momentum $\hbar q_n$ that are equal to integer multiples of $\hbar q_c=\hbar\pi /L$ corresponding to a  critical stirring velocity $v_c=\pi \hbar /mL$. This result shows that it is difficult to stir the mesoscopic TG gas except at velocities that are multiples of $v_c$. At generic values of the stirring momentum away from $\hbar q_n$ the fluid belongs to the zero current branch, which for nonadiabatic stirring momenta $\hbar q>\hbar q_c$ corresponds to an excited state. The occupation of such highly excited states is due to the instantaneous switching on of the barrier motion. At the special values of the stirring momentum $\hbar q_n$ avoided crossings between single-particle branches with different angular momentum occur. This allows for the occurrence of superpositions of states with different angular momentum \cite{SchMinHek11} which yield a non zero average current. In contrast to the nonadiabatic stirring mechanism, an adiabatic switching on leaves the system in the ground manifold $\left\{k_j\right\}$ with $j=1...N$ and corresponding energy $E_g(q)=\sum_{j=1}^N \hbar^2 k_j^2/2m$. In this case the integrated, time averaged  current is given by the thermodynamic relation $I=(N\hbar q/m-\partial E_g/\partial \hbar q)/L$  and has a staircase behavior as a function of the stirring momentum, also illustrated in Fig.\ref{fig2}a. 
 
Figure \ref{fig2}a shows for reference the corresponding time averaged current of an ideal Bose gas subjected to the same stirring drive. For momenta close to $\hbar q_c$  we find that the  current induced in an ideal Bose gas is larger than the one of the strongly interacting TG gas. This apparently surprising result is readily explained by noticing that the ideal Bose gas occupies the lowest single-particle level which is the most affected by a weak delta-barrier potential since $u_q^{(1)}$ and $v_q^{(1)}$ show a very smooth dependence on $q$. For large stirring momenta $\hbar q\gg\hbar q_c$ this effect vanishes since then also the ideal Bose gas is in a very highly excited state. In the inset of Fig.\ref{fig2}a we show that a very good agreement is found at weak barrier strength between perturbation theory and exact calculation for momenta close to $\hbar q_n$ .

Finally, we remark that our results for the integrated current are also in agreement with the prediction of nonsuperfluid behavior of a TG gas in an infinitely long wire. In that case, $L$ tends to infinity keeping $N/L$ constant, and the critical velocity $v_c$ tends to zero. The observed superfluid effect is truly mesoscopic, i.e. associated with the finite size of the ring.

\begin{figure}[t]
\includegraphics[width=0.40\textwidth]{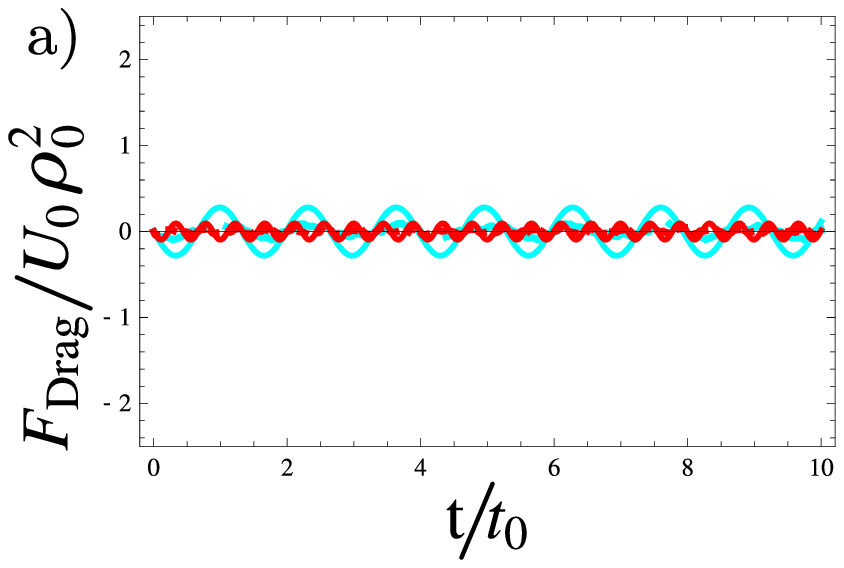}
\includegraphics[width=0.40\textwidth]{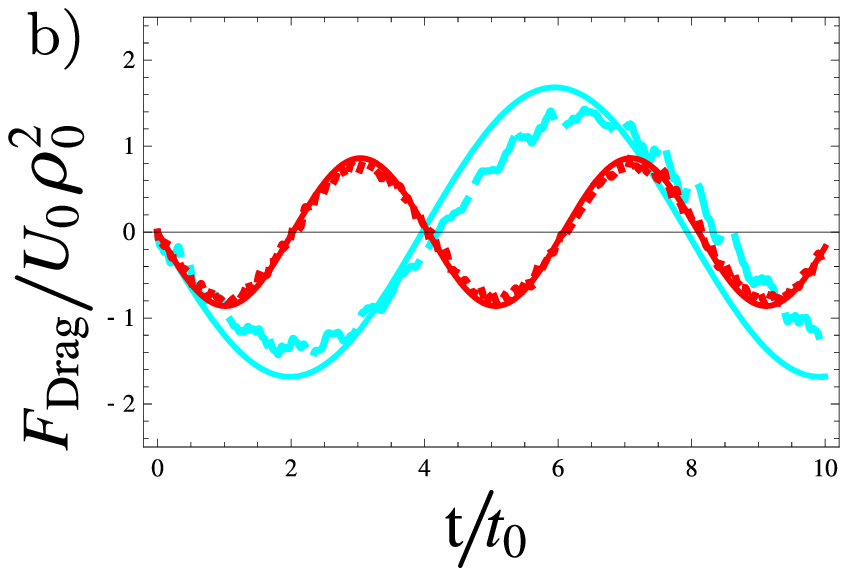}
\includegraphics[width=0.40\textwidth]{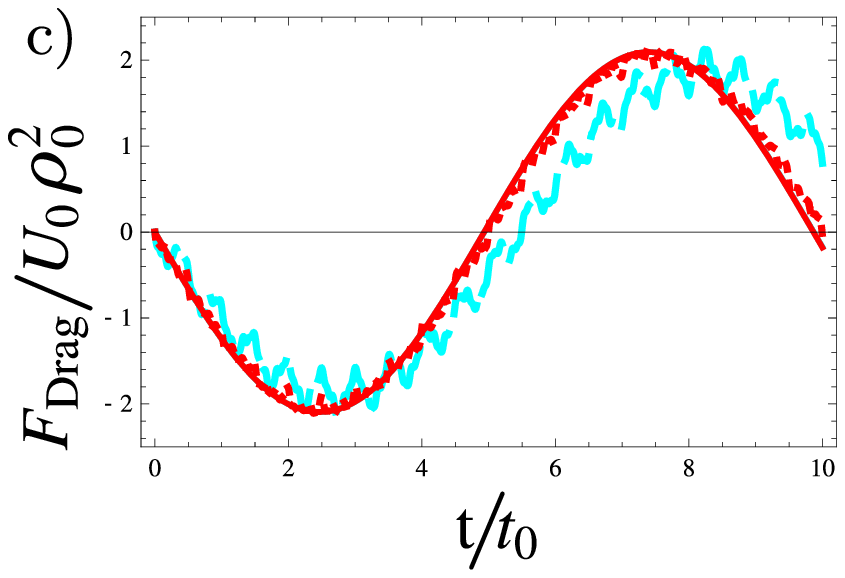}
\caption{(Color online) The drag force exerted by the fluid on the moving barrier in units of $U_0 \rho_0^2$, $\rho_0=N/L$ being the average density of the fluid, as a function of time in units of $t_0=mL^2/\pi\hbar$ for a TG gas (red dotted curves) and an ideal Bose gas (cyan dashed curves) for three values of stirring wavevector, $q=0.25$ $q_c$, (a), $q=0.925$ $q_c$(b) and $q=q_c$ (c), for $N=3$ particles and a barrier strength  $m U_0L/\hbar^2=1$. The numerical results are compared with the results of the perturbative approach (thin solid lines). In panel (c) for $q=q_c$ the analytical curves for ideal bosons and the TG gas are identical since $\delta q=0$ and therefore $\Delta E_q^{(l)}$ and $\left\{u_q^{(l)},v_q^{(l)}\right\}$ are independent of $l$.}
\label{fig3}
\end{figure}

\section{Particle current fluctuations}

To further explore superfluid behavior, we complement the study of the integrated particle current by analyzing its fluctuations. This quantity measures how well defined the particle current is with respect to unavoidable quantum fluctuations. The time-averaged particle current fluctuations are defined as 
\begin{equation}
\label{Eq:fluctuations}
\Delta I=\sqrt{\langle I^2\rangle-\langle I\, \rangle^2}\ ,
\end{equation}
where the time-averaged integrated particle current is given by $\langle I\, \rangle=1/T\int dt I(t)$. The second moment of particle current reads
\begin{equation}
\label{Eq:currentcurrentInt}
\langle I^2\rangle=\frac{1}{L^2}\frac{1}{T}\int_{-T/2}^{T/2}\int^{L/2}_{-L/2} dx dy\ C_2(x,y,t)\ ,
\end{equation}
where 
\begin{equation}
\label{Eq:currentcurrent}
C_2(x,y,t)= \langle \Psi_B|j(x,t)j(y,t)|\Psi_B\rangle\ 
\end{equation}
is the spatially integrated equal-times current-current correlation function.

The particle current fluctuations derived from perturbation theory read
\begin{multline}
\label{eq:int_currentcurrentPT}
\Delta I= \left(\frac{2\hbar}{mL}\right)^2\! \sum_{l=1}^N 2(q_n-k_l)^2\left(u_q^{(f_{nl})}v_q^{(f_{nl})}\right)^2\\ \times\left[1-3\left(u_q^{(f_{nl})}v_q^{(f_{nl})}\right)^2\right]\, .
\end{multline}
Fig.\ref{fig2}b shows the time-averaged particle current fluctuations for a non-adiabatic stirring. For stirring momenta far away from $\hbar q_n$ the qualitative behavior of the fluctuations is similar to the one of the current itself. Consequently in the superfluid regime, away from $q_n$ there is neither an induced current nor strong current fluctuations. The quantum state of the system corresponds to a well-defined angular momentum state which explains the smallness of the fluctuations. On the other hand, around the current peaks, i.e. for stirring momenta $\hbar q_n$, we observe strong fluctuations. At these special points, the system oscillates between two equally possible angular momentum states, leading to the appearance of strong current fluctuations. At $q_n$ we further observe the formation of a dip in the fluctuations, which can be understood qualitatively as follows. The nonadiabatic current in perturbation theory is given by the weight combination $u_q^{(l)}v_q^{(l)}$, a structure similar to the \textit{second} cumulant characteristic for the partition noise for particles tunneling across a barrier \cite{BlanBuett00}. Its fluctuations are consequently given by the \textit{fourth} cumulant of the same distribution which gives these splitted peak structures.

\section{Drag force}
The drag force \cite{astrakharchik_stirring} acting on a barrier potential $V_{barr}(x)$ is defined as $F_d=-\langle \partial_x V_{barr}(x) \rangle$ where $\langle... \rangle$ denotes the quantum average with respect to  the state of the system.
For the specific case of  a delta barrier potential $U_0\delta(x)$ the drag force is simply related to the particle density profile $n(x)$, according to 
\begin{equation}
\label{Eq:drag}
F_d=U_0 \frac{1}{2} [\partial_x  n(x)|_{x=0^+}+\partial_x  n(x)|_{x=0^-}].
\end{equation}
For a TG gas the density profile is readily calculated within our exactly solvable model, as it coincides with the one of a noninteracting Fermi gas,
\begin{equation}
\label{Eq:densityprof}
n(x,t)=\sum_{l=1}^N |\psi_l(x,t)|^2.
\end{equation}
The expression for the drag force acting on a delta barrier potential suddenly set into motion across a TG gas in perturbation theory reads
\begin{multline}
\label{eq:int_DragPT}
F_d= \frac{4U_0}{L} \sum_{l=1}^N (q_0-k_l)\left(u_q^{(f_{nl})}v_q^{(f_{nl})}\right)^2\sin(2\Delta E_q^{(f_{nl})}t)\, .
\end{multline}

Fig.\ref{fig3} shows the drag force obtained from both the exact model Eqs.(\ref{Eq:drag}), (\ref{Eq:densityprof}) and (\ref{eq:orbitals}) and from perturbation expansion at weak barrier strength. It is an oscillating function of time with a typical frequency associated with the energy-eigenvalues separations $\Delta E_q^{(l)}$ whose main contribution comes from the Fermi level, $E_{N+1,q}-E_{N,q}$. Notice that since the energy levels depend on the stirring momentum the typical oscillation frequency is also momentum dependent. At increasing stirring momentum, we find that the behavior of the drag force closely follows the results for the integrated current presented in the previous section.
In the case of $\hbar q<\hbar q_c$ the drag force is vanishingly small, confirming the picture of a superfluid mesoscopic TG gas. If the barrier momentum is instead chosen  close  or equal to $\hbar q_c$, then it is possible to transfer angular momentum to the fluid and the drag force is nonvanishing, illustrating the breakdown of superfluidity.  We find the same velocity threshold as the one predicted in Ref.\cite{buchler} using a Luttinger-liquid description of a weakly interacting Bose gas.   Finally, for momenta larger than $\hbar q_c$  except at the special momenta close to integer multiples of $\hbar q_c$, we find that the drag force vanishes as the system is in a very peculiar, highly excited state which does not couple to the barrier motion.

In Fig.\ref{fig3} we also show the drag force originating from the  stirring of an ideal Bose gas as a function of time. The typical oscillation frequency of the drag force of an ideal Bose gas is fixed by the first energy level splitting $E_{2,q}-E_{1,q}$. Due to the flatter dispersion  of the first energy levels   as compared to those at the Fermi surface, for the ideal Bose gas the  dependence of the drag-force  oscillation frequency on the velocity is weaker than the one found for the TG gas. The three panels of Fig.\ref{fig3} correspond to increasing values of the stirring momentum. At small momenta also the ideal Bose gas shows mesoscopic superfluidity. For momenta close to $\hbar q_c$ the drag force of the ideal Bose gas is larger in magnitude than the one of the TG gas, illustrating the more important effect of the barrier on the ideal Bose gas. For a stirring momentum equal to $\hbar q_c$ the drag force of the TG gas and of the ideal Bose gas are equal in magnitude, as it is the case of the integrated current (see again Fig.\ref{fig2}a).

\section{Conclusions}
In summary, using an exactly solvable model, we have investigated the transport properties of a TG gas on a ring trap stirred by a localized barrier potential moving along the ring circumference. The calculation of the spatially integrated particle current its fluctuations and of the drag force acting on the barrier shows the existence of a velocity threshold for setting the TG gas into motion, at a value given by $v_c=\hbar \pi/mL$. The same  value has been predicted for the weakly interacting mesoscopic Bose gas by a Luttinger-liquid calculation \cite {buchler}. We have also shown that depending on the way the final velocity of the barrier is reached, different quantum  states are accessed. In particular, we have found  that if the barrier motion is suddenly switched on, it is difficult to transfer angular momentum to the system except close to $v_c$ or its integer multiples. Finally, by comparing with the corresponding results for a noninteracting 1D Bose gas, we have shown that the stirred current, the current fluctuations and the drag force depend on the interaction strength. These limiting cases allow us to obtain a qualitative picture for the behavior of Bose gases for both very weak and very strong interactions. Given the experimental progress in recent years, the realization of a experimental set up that allows to study the mesoscopic TG gas on a ring seems close to reach. In perspective, it would be interesting to develop tools for describing the case of a 1D Bose gas with arbitrary interactions subjected to an out of equilibrium drive.

\begin{acknowledgments} 
We acknowledge discussions with Denis Basko, Edouard Boulat, Roberta Citro and Helene Perrin. We thank the PEPS-PTI project ``Quantum gases and condensed matter'', the MIDAS STREP project and the Handy-Q ERC project for financial support. 
\end{acknowledgments}


\begin{thebibliography}{10}
\bibitem{note}
For even $n$ the first band $j=1$ is in a definite state of angular momentum and its wavefunction reads $\phi_{q}^{(1)}(x)=\exp(-iqx)\exp(i \pi n x/L)/\sqrt{L}$.
\end{thebibliography}
\end{document}